\documentclass[a4paper,11pt]{article}
\pdfoutput=1 
\usepackage{jinstpub} 
\usepackage{float}

\usepackage{caption}
\usepackage{subcaption}

\title{\boldmath Optimization on fixed low latency implementation of the GBT core in FPGA}

\author{Kai Chen\thanks{Corresponding
		author.},~ Hucheng Chen, Weihao Wu, Hao Xu and Lin Yao}
\affiliation{Brookhaven National Laboratory,\\
	PO Box 5000, Upton, NY 11973, USA}	

\emailAdd{kchen@bnl.gov}

\abstract{In the upgrade of ATLAS experiment\,\cite{bibATLAS}, the front-end electronics components are subjected to a large radiation background. Meanwhile high speed optical links are required for the data transmission between the on-detector and off-detector electronics. The GBT architecture and the Versatile Link (VL) project are designed by CERN to support the 4.8 Gbps line rate bidirectional high-speed data transmission which is called GBT link\,\cite{bibGBT}. In the ATLAS upgrade, besides the link with on-detector, the GBT link is also used between different off-detector systems. The GBTX ASIC is designed for the on-detector front-end\,\cite{bibGBTX}, correspondingly for the off-detector electronics, the GBT architecture is implemented in Field Programmable Gate Arrays (FPGA). CERN launches the GBT-FPGA project to provide examples in different types of FPGA\,\cite{bibGBTFPGA}. In the ATLAS upgrade framework, the Front-End LInk eXchange (FELIX) system\,\cite{bibFELIX,bibTWEPP2016} is used to interface the front-end electronics of several ATLAS subsystems. The GBT link is used between them, to transfer the detector data and the timing, trigger, control and monitoring information. The trigger signal distributed in the down-link from FELIX to the front-end requires a fixed and low latency. In this paper, several optimizations on the GBT-FPGA IP core are introduced, to achieve a lower fixed latency. For FELIX, a common firmware will be used to interface different front-ends with support of both GBT modes: the forward error correction mode and the wide mode. The modified GBT-FPGA core has the ability to switch between the GBT modes without FPGA reprogramming. The system clock distribution of the multi-channel FELIX firmware is also discussed in this paper.
}

\keywords{Fixed low latency transmission; GBT link; Transceiver in FPGA}

\begin{document}


\maketitle	

\section{Introduction}
The ATLAS Phase-I upgrade requires a Trigger and Data Acquisition (TDAQ) system able to trigger and record data from up to three times the nominal LHC instantaneous luminosity. The new Front-End LInk eXchange (FELIX) system provides a scalable PC-based gateway to interface custom radiation tolerant optical links from front-end electronics, via PCIe Gen3 cards, to a commercial switched Ethernet or InfiniBand network\,\cite{bibFELIX,bibTWEPP2016}. 
As a system, FELIX consists of a PC running a Linux based OS (SLC6), an Ethernet or InfiniBand Network Interface Card (NIC), and up to two FELIX PCIe cards. Besides the PCIe cards which interface the front-ends, all other parts will be commercial. In Phase-I upgrade, it will interface several ATLAS subsystems. This includes the Muon new small wheel for muon trigger, the LAr (Liquid Argon) Calorimeter trigger readout electronics, and the L1Calo (Level-1 Calorimeter Trigger) trigger electronics upgrade for electron trigger. The down link from FELIX to front-ends will be the 4.8 Gbps GigaBit Transceiver (GBT) link designed by CERN, the up link will be the GBT link or the 9.6 Gbps full mode link developed by the FELIX group\,\cite{bibTWEPP2016}. The full mode link is only used by FPGA based front-ends.

The GBT architecture is a chipset designed by CERN. It provides a method to increase the data transmission bandwidth and has the ability to sustain high radiation dose. As an on-detector radiation-hard ASIC, the GBTX ASIC is part of the GBT chipset\,\cite{bibGBTX}. It supports 4.8 Gbps bidirectional data transfer with the off-detector via the Versatile Link designed at CERN. The data port of the GBT core is 120 bits in 40 MHz clock domain. This 120-bit GBT frame consists of 116-bit data and 4-bit header. The 4-bit header is the flag for the data alignment at the receiver side, it can only be data (0101) or idle (0110) in each frame. Four of the 116-bit data is used for the slow control to front-end, based on the High-Level Data Link Control (HDLC) protocol\,\cite{bibHDLC}. There are two encoding modes for the 116-bit of data, the low overhead wide mode and normal Forword Error Correction (FEC) mode. For the wide mode, four 21-bit scramblers and two 16-bit scramblers will scramble the 116-bit input data. After the scrambling, the 120-bit data is serialized and sent via the optical fiber. For the FEC mode, four 21-bit scramblers are used to scramble the 84-bit input. Two FEC16 modules are implemented to generate the 32-bit FEC field from the scramblers' output. The FEC supports correction of up to 16 consecutive bits of error\,\cite{bibGBTX}. At the receiver side, the 120-bit received data is decoded and descrambled. The link consists of the GBTX ASIC at the front-end side and the matching GBT interface at the back-end, which is realized in FPGAs. CERN's GBT group launches the GBT-FPGA project to supply the GBT firmware example for different kinds of FPGA\,\cite{bibGBTFPGA,bibGBTLAT}. In the CERN GBT-FPGA core, the scrambling, descrambling, FEC encoding and decoding blocks are all in the 40 MHz TTC (Timing, Trigger and Control) clock domain.

The GBT core is implemented in the FELIX firmware, to communicate with the GBTX ASIC or the front-end FPGA containing a GBT core. The TTC signals\,\cite{bibTTC}, and the front-end specified control commands are transferred from the FELIX via the down link to the front-end. The monitoring information and other detector data are transmitted via the up-link from the front-end to FELIX. The Phase-I upgrade of the ATLAS experiment is aiming to improve the trigger selectivity, particularly the single lepton trigger. The Level-1 trigger latency of the ATLAS experiment is fixed (2.5us), therefore the latency reduction along the trigger electronics path becomes critical. The low latency is crucial for both trigger generation path: from detector front-ends to the CTP (central trigger processor), and trigger distribution path: from the CTP to detector front-ends. For FELIX in the Phase-I upgrade, the fixed low-latency is required for the distribution of trigger signals through the down-links. 

The CERN GBT-FPGA core is provided with two versions: the low-latency version and the standard latency version. Reference\,\cite{bibGBTLAT} introduces some effects to decrease the latency. In our design, an optimized version was built up based on kernel modules of the CERN GBT core. Some optimizations are applied with the purpose to decrease latency at both of the transmitter (TX) and receiver (RX) sides. FELIX firmware needs to support interfacing different front-ends with different GBT modes. The modified GBT-FPGA core has the ability to switch the GBT mode at both of the TX and RX sides without FPGA firmware reloading. Meanwhile one FELIX card will support multiple GBT links. The details about the optimizations and multi-channel design will be presented in this paper.

\section{Optimizations to the GBT-FPGA core}

\begin{figure}[H]
	\centering
	{
		\includegraphics[width=0.8\textwidth]{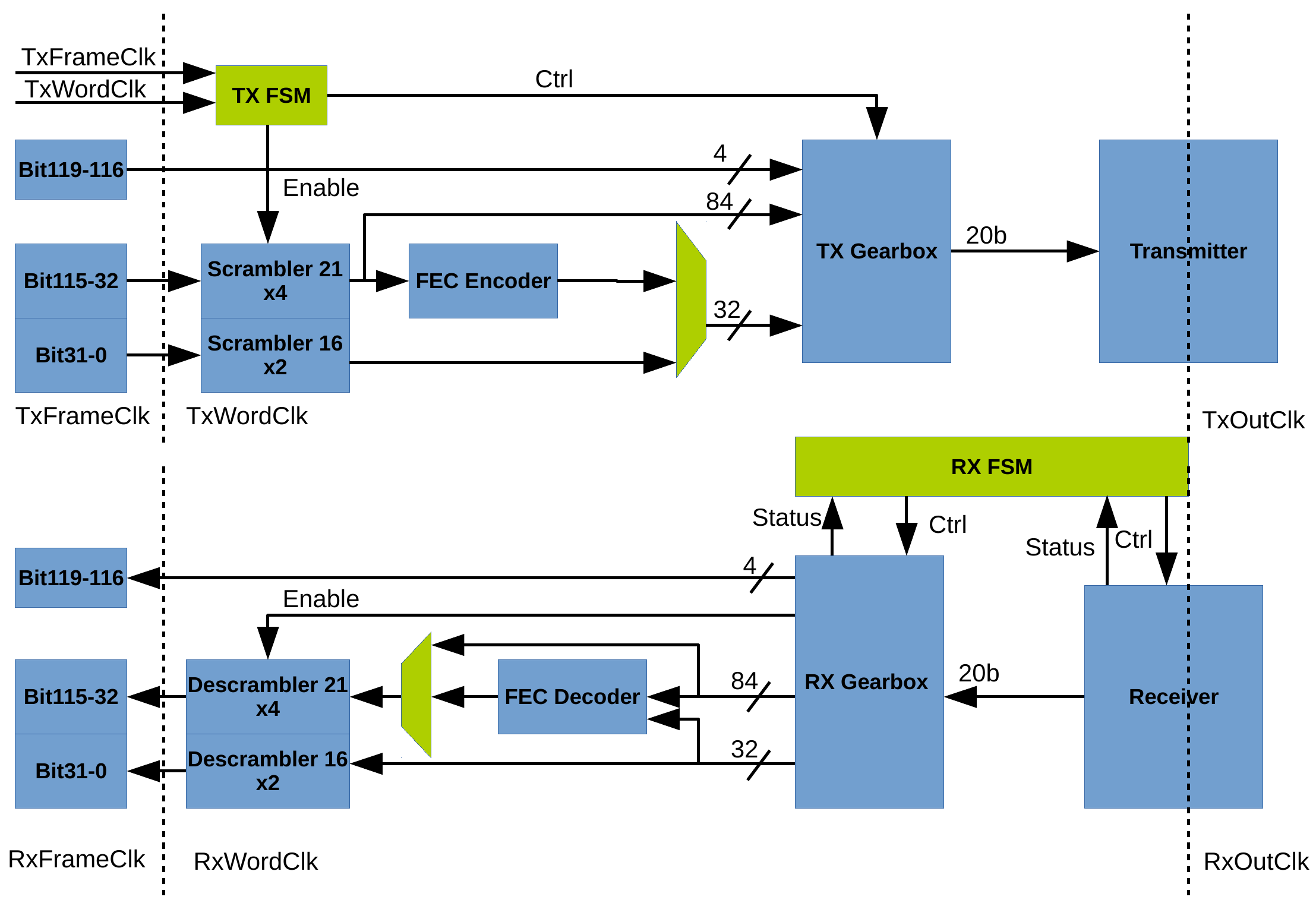}}
	\caption{Block diagram of the modified GBT core}\label{changes}
\end{figure}

The kernel modules of the GBT are the FEC encoder, FEC decoder, scramblers and descramblers with width of 16 and 21. As shown in Figure\,\ref{changes}, after the encoding, the 120-bit data needs to be serialized. Since the data port of the transceiver is 20 bits wide, a TxGearbox is needed to shift the 120-bit data into the transmitter in 6 cycles of TxWordClk. Similarly, the RxGearbox is used in the receiver side. Two multiplexers are added as depicted in Figure\,\ref{changes}. At the TX side, the lower 32-bit FEC field to the TxGearbox can be configured as from the output of the FEC encoder, or from the two 16-bit scramblers. At the RX side, the FEC decoder can be bypassed for the wide mode. By configuring these two multiplexers, the GBT mode for the TX and RX can be dynamically changed separately, without reloading a different FPGA firmware, which is different from the CERN GBT core. With the design in Figure\,\ref{changes}, all GBT encoding and decoding is processed in 240 MHz clock domains, to get a lower latency for both of the TX and RX sides. To benchmark the latency, several time units are used: 1 UI (Unit Interval) means 1/4.8GHz $\approx$ 208.3 ps; 1 Cycle means 1/240MHz $\approx$ 4.167 ns; 1 BC (Bunch Crossing) means 1/40MHz = 25 ns.

\subsection{The transmitter side}
\begin{wrapfigure}[10]{r}{0.57\textwidth}
	\vspace{-3mm}
	\centering
	\includegraphics[width=0.57\textwidth]{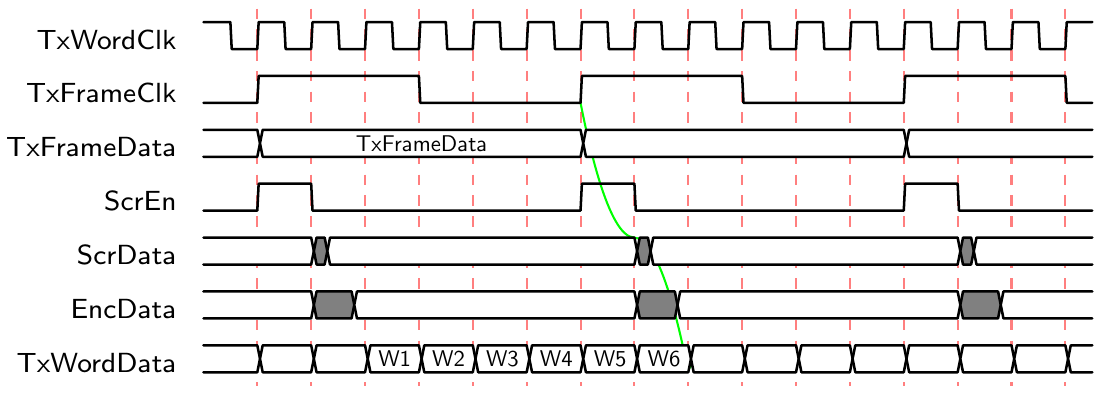}
	\vspace{-2mm}
	\caption{The timing diagram of the GBT encoding: the arrow is from the GBT encoding start point to stop point}
	\label{txtd}
\end{wrapfigure}

At the transmitter side, typically the 120-bit GBT data needs to be transferred from the 40 MHz TxFrameClk to the 240 MHz TxWordClk. For the ATLAS experiment, the TxFrameClk is synchronized with the TTC clock. The CERN GBT core places the time domain crossing before the TxGearbox. To decrease the latency, this clock domain crossing is moved to before the scramblers. Then all of the GBT encoding blocks operate in the domain of the 240 MHz TxWordClk. Being comprised of only some XOR gates, the scrambling operation can be finished in one cycle of TxWordClk easily, but after the change of domain, the scrambling should only be enabled in 1 of 6 Cycles. The scramblers are modified to be controllable by the control signal from the module TX\_FSM (TX Finite State Machine). As shown in Figure\,\ref{txtd}, the TxWordClk samples the TxFrameClk to find in which cycle the rising edge is, and then the control signal ScrEn for scramblers is generated. This signal also determines in which cycle the TxWordClk samples the GBT frame data (TxFrameData) transferred from the 40 MHz TxFrameClk domain. In the TX\_FSM, the ScrEn can be adjusted in steps of 1 Cycle. By changing it, a low latency with enough setup time for the clock domain crossing can be obtained. The ScrData and Encdata in this figure are data before and after FEC encoding. As shown in Figure\,\ref{changes}, the TX\_FSM also generates a control signal for the TxGearbox module. It is used to start shifting the 120-bit data from the second cycle after the ScrEn pulse, the first word to send is W1 in Figure\,\ref{txtd}. For the FEC mode, the encoding operation needs extra time, but it only affects the 32-bit FEC field. Since these 32 bits are in the last two words of W5 and W6, the transmission of W1-W4 doesn't need to wait until FEC encoding is finished.

\begin{wrapfigure}[10]{r}{0.57\textwidth}
	\vspace{-3mm}
	\centering
	\includegraphics[width=0.57\textwidth]{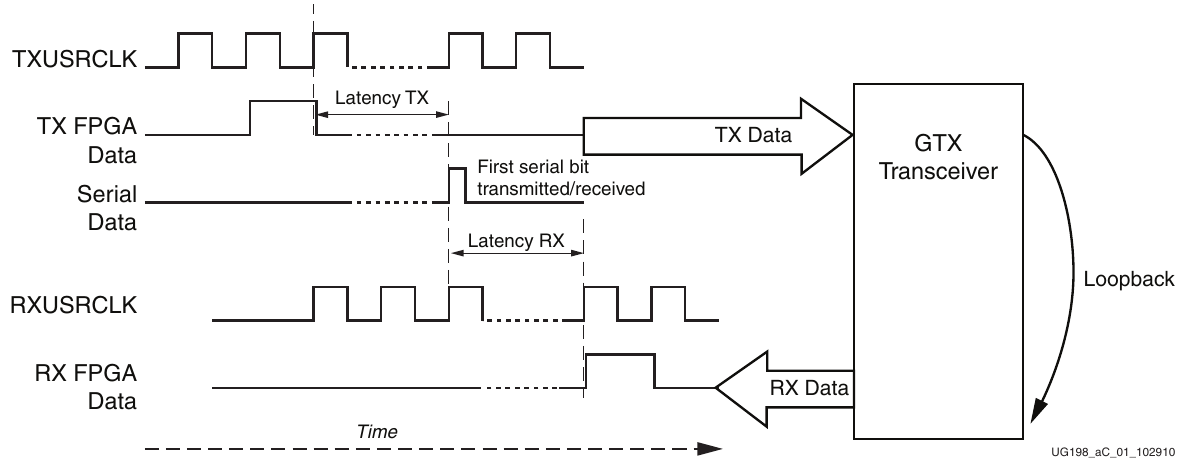}
	\vspace{-2mm}
	\caption{Definition of latency for the transceiver\,\cite{bibXILINXV6}}\label{latdef}
\end{wrapfigure}

The start point at TX side and the stop points at RX side should be defined to calculate and compare the total latency. To make the latency value comparable, the same definition with the CERN example design is used. The TX side latency is split into two parts: the GBT encoding and the FPGA transmitter. For the GBT encoding, as seen in Figure\,\ref{txtd}, the start point is the rising edge of the TxFrameClk aligned to the end of the labeled Tx frame data. The stop point is when the 20-bit TxWordData sent to the transceiver changes from W6 to W1 of the next frame. The latency value is 1$\sim$2 cycles, it is deterministic and determined by the phase between the TxWordClk and TxFrameClk. As defined in the Xilinx manual\,\cite{bibXILINXV6}, the end point of the whole TX side is when the first serial bit of the last word W6 is transmitted out from the FPGA transmitter, as shown in Figure\,\ref{latdef}. This design is tested on a Xilinx KC705 evaluation board\,\cite{bibKC705}, based on the calculation with parameters taken from Xilinx datasheets\,\cite{bibXILINXK7}, the latency of GTX transmitter is: 
\begin{equation}
	LAT_{TxGTX}=4\cdot Cycle + 33.73125\cdot UI= 23.7\ ns
\end{equation}
Totally, the latency of TX side is:
\begin{equation}
	LAT_{TX}=(1\sim2)\cdot Cycle + (4\cdot Cycle + 33.73125\cdot UI)= 27.8\sim32\ ns
\end{equation}

\subsection{The receiver side}
At the receiver side, typically the GBT frame data needs to be transferred to the 40 MHz RxFrameClk from the 240 MHz RxWordClk as shown in Figure\,\ref{changes}. Both of these two clocks are recovered from the data stream on the GBT link. The domain crossing is done after the RxGearbox module in the CERN GBT core. To decrease the latency, the descramblers and FEC decoder are moved into the 240 MHz RxWordClk domain. The clock domain crossing is after the descrambler, the 120-bit descrambler output RxFrameData can also be transferred to a clock domain like 80, 160 or 320 MHz. We can generate these clocks from the recovered 40 MHz RxFrameClk, resulting in fixed phases.

\begin{wrapfigure}[12]{r}{0.55\textwidth}
	\vspace{-3mm}
	\centering
	\includegraphics[width=0.55\textwidth]{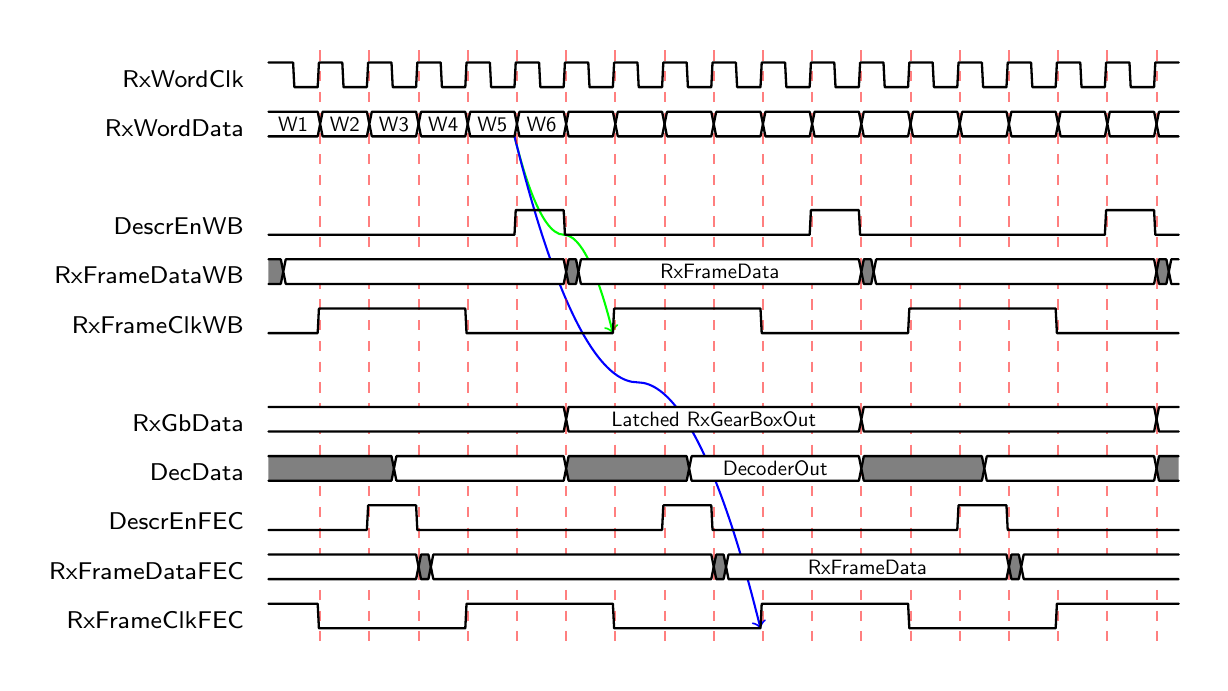}
	\vspace{-2mm}
	\caption{The timing diagram for the GBT decoding: the arrows are from the GBT decoding start point to stop points, for the two GBT modes}
	\label{rxtd}
\end{wrapfigure}

The FSM module RX\_FSM in Figure\,\ref{changes} will do the receiver reset and bitslip to lock the GBT header at the expected bit 3-0 of the RxWordData in Figure\,\ref{rxtd}. The details about the bitslip and header locking will be introduced in Chapter\,\ref{chap:4}. A six value counter is run inside the RxGearbox. Indicated in Figure\,\ref{rxtd}, the RxWordData with the GBT header inside is W1. It will determine phase of the six value counter and generate the enable signal DescrEn for the descramblers based on the counter value. Silimar to the scramblers, the descramblers are also modified to support the external control. Different from the TX side, the time to pull up the DescrEn signal will change with the GBT mode. For the wide mode, the 20-bit data RxWordData from the transceiver is not latched, the descrambling process can start at the edge when the data changes from W6 to W1 of the next frame. For the FEC mode, the RxGearbox output must be latched since the FEC decoding operation can not finish in one RxWordClk cycle. The following FEC decoding module will use the latched RxGearbox output data (RxGbData in Figure\,\ref{rxtd}). The FEC decoding needs about 2.5 Cycles in different FPGAs we have tested, so the DescrEn of the FEC mode is set to be 3 cycles later than that of the wide mode. From the Figure\,\ref{rxtd}, we can see when the DescrEnFEC is enabled, the FEC decoder output DecData is already stable. To set the timing constraint for the FEC decoder, the multi-cycle path constraint provided by Xilinx is applied for the data transferred from the RxGearbox to the descramblers. 

The receiver side latency also consists of two parts: the FPGA receiver and the GBT decoding module. The start point of the FPGA receiver is when the first serial bit of W6 is received. The stop point is when the 20-bit parallel data RxWordData changes from W5 to W6\,\cite{bibXILINXV6}. Same with the CERN version latency definition, the stop point of the GBT decoding is when the GBT frame data is sampled by the recovered 40MHz RxFrameClk. For the FPGA on the KC705 board, the calculated latency for the GTX receiver is about:
\begin{equation}
LAT_{RxGTX}=3\cdot Cycle + 110.5\cdot UI= 35.52\ ns
\end{equation}
Totally, for the wide mode, latency of the RX side is:
\begin{equation}
LAT_{RxWB}=2\cdot Cycle + (3\cdot Cycle + 110.5\cdot UI)= 43.9\ ns
\end{equation}
For the FEC mode, the latency will be:
\begin{equation}
LAT_{RxFEC}=5\cdot Cycle + (3\cdot Cycle + 110.5\cdot UI)= 56.4\ ns
\end{equation}
Since the dynamic mode changing is supported, to simplify the design, we let the wide mode also uses the latched RxGbData as the descramblers input. That means the descrambler enable signal DescrEnWB, the recovered data and the clock are all delayed by 1 Cycle for the wide mode. Then:
\begin{equation}
LAT_{RxWB}=3\cdot Cycle + (3\cdot Cycle + 110.5\cdot UI)= 48.0\ ns
\end{equation}

\section{Test result on the KC705 board}
From the above calculation, the latency of the whole data path is:
\begin{equation}
LAT_{WB}=(27.8\sim32.0) + 48.0 = 75.8\sim80.0\ ns
\end{equation}
\begin{equation}
LAT_{FEC}=(27.8\sim32.0) + 56.4 = 84.2\sim88.4\ ns
\end{equation}

\begin{wrapfigure}[13]{r}{0.57\textwidth}
	\vspace{-3mm}
	\centering
	\includegraphics[width=0.57\textwidth]{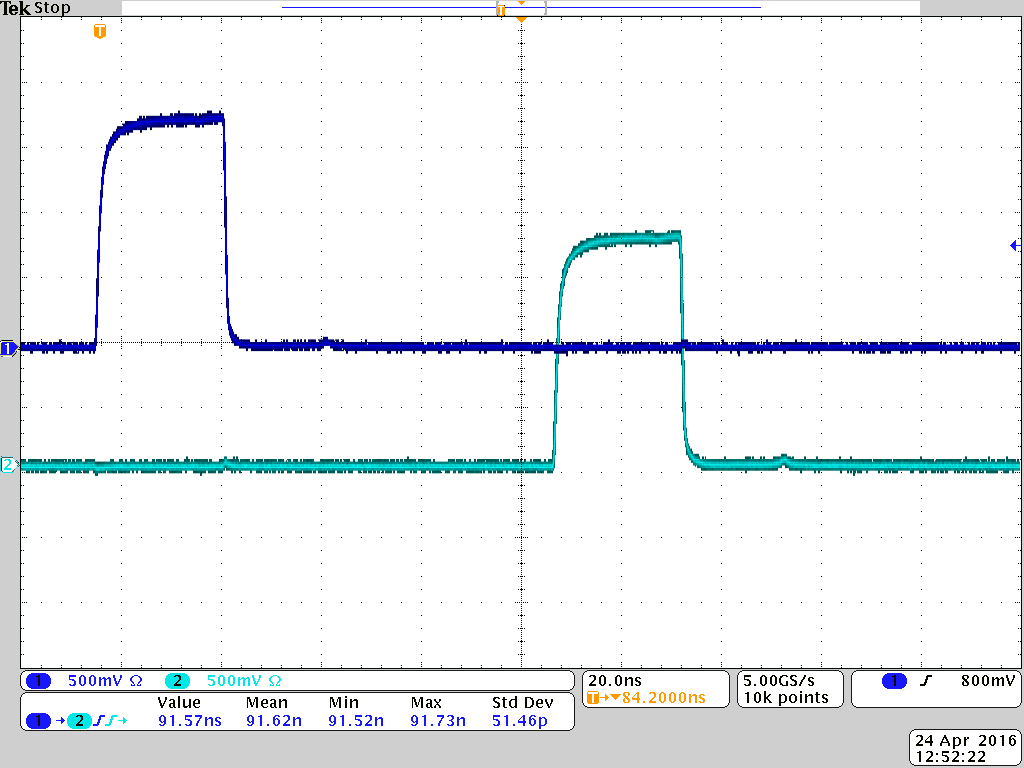}
	\vspace{-2mm}
	\caption{The latency test on KC705 for FEC mode}\label{testkc705}
\end{wrapfigure}

The latency between two KC705 boards connected by one meter fiber is measured. The same data bit at the TX and RX side are connected to a scope via two SMA connectors. The result of the FEC mode is shown in Figure\,\ref{testkc705}. The latency measured by the scope is 91.6 ns.  Excluding the 5 ns latency of the one meter fiber, the latency is about 86.6 ns. The standard deviation of the latency measurement is about 50 ps. With the wide mode, the latency is about 78.3 ns. For the CERN GBT-FPGA example on KC705, the whole latency with the low-latency configuration is about 133.8 ns. Totally the optimization will save about 2 BC. The comparison of latency results is shown in Table\,\ref{tab:kc705}. Since no additional module is added, the latency optimization doesn't require more logic resource. Regarding the timing, the FEC decoder needs about 2.5 RxWordClk cycles. In our firmware, the time margin kept for it can be adjusted in the step of 1 Cycle, to come to a compromise between the timing tolerance and the latency requirement. A margin with more cycles means bigger latency, but it alleviates the requirement to the firmware running speed. This is useful when the FPGA resource utilization is in a high percentage, or when a slow FPGA is used.

\begin {table}[H]
\begin{center}
	\begin{tabular}{ | l | r | r | r |}
		\hline
		& TX (ns) & RX (ns) & TX+RX (ns) \\ \hline\hline
		Virtex 6 (CERN version) & 47.9\protect\footnotemark & 82.4 & 130.3  \\ \hline
		KC705 testing (CERN version) &  &  & 133.8  \\ \hline
		KC705 calculation: Wide mode & 27.8$\sim$32 & 43.9 & 71.7 \\ 
		\hline
		KC705 calculation: FEC mode  & 27.8$\sim$32 & 56.4 & 84.2 \\ 
		\hline
		KC705 calculation: on-line mode switch & 27.8$\sim$32 & 48.0/56.4 & 75.8$\sim$80.0/84.2$\sim$88.4 \\ 
		\hline
		KC705 testing: on-line mode switch &  &  & 78.3/86.6 \\
		\hline
	\end{tabular}
	\caption {Results of the testing on KC705 board}\label{tab:kc705} 
\end{center}
\end {table}
\addtocounter{footnote}{0}
\footnotetext{Definition of the boundary between TX and RX may be different with Figure\,\ref{latdef}.}

This optimized GBT-FPGA core was run for more than 60 hours. No bit error occurred during this long time test, so the bit error rate is smaller than 10$^{-15}$. The latency was stable even after power cycling and resetting the transceiver, this meets the requirement of the FELIX system.

\section{Clock sharing in the multi-channel design}\label{chap:4}

\begin{wrapfigure}[23]{r}{0.42\textwidth}
	\vspace{-3mm}
	\centering
	\includegraphics[width=0.42\textwidth]{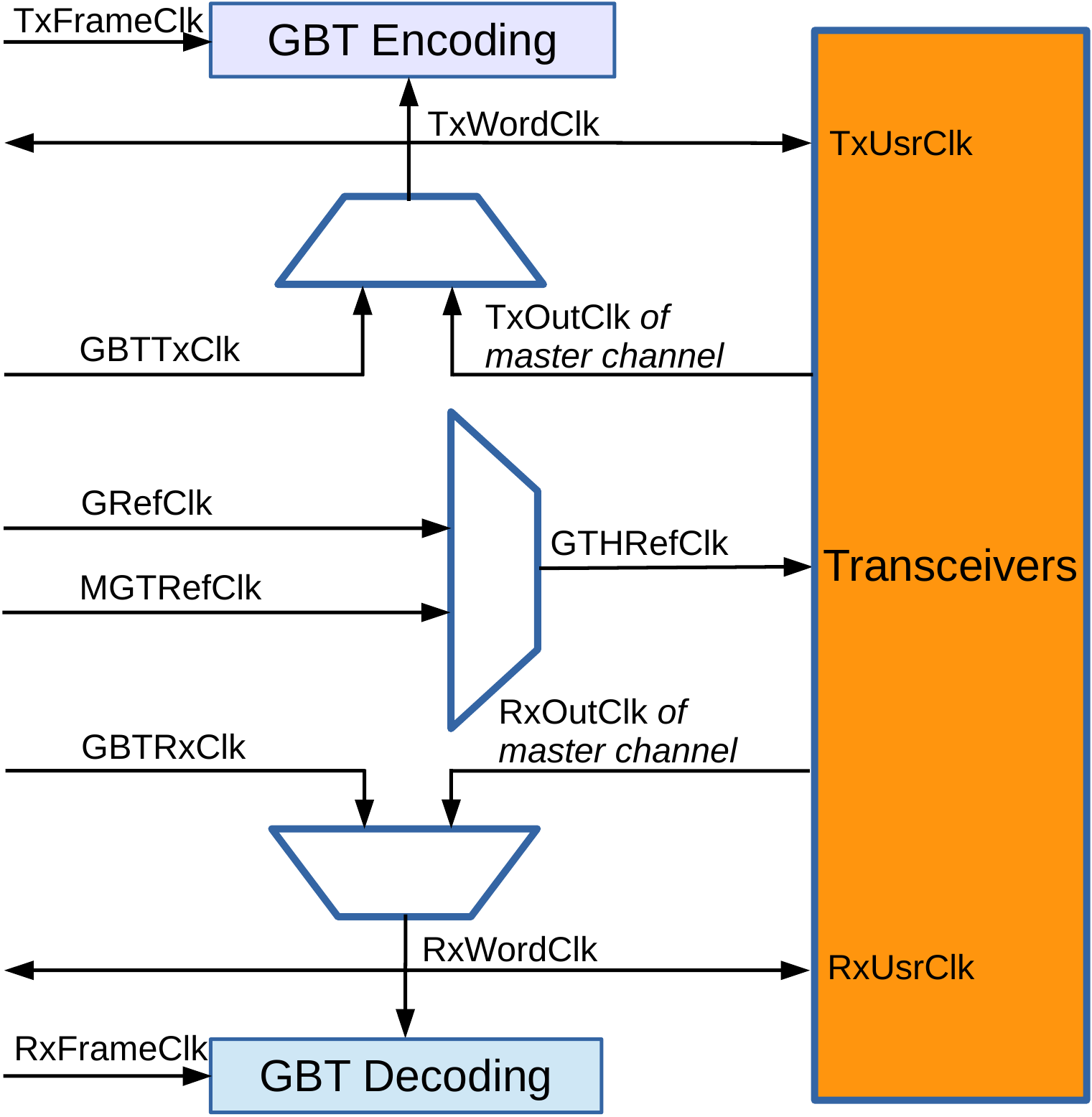}
	\vspace{-2mm}
	\caption{Clock distribution of multi-channel design}\label{clk_dist}
	\vspace{0mm}
	\centering
	\includegraphics[width=0.42\textwidth]{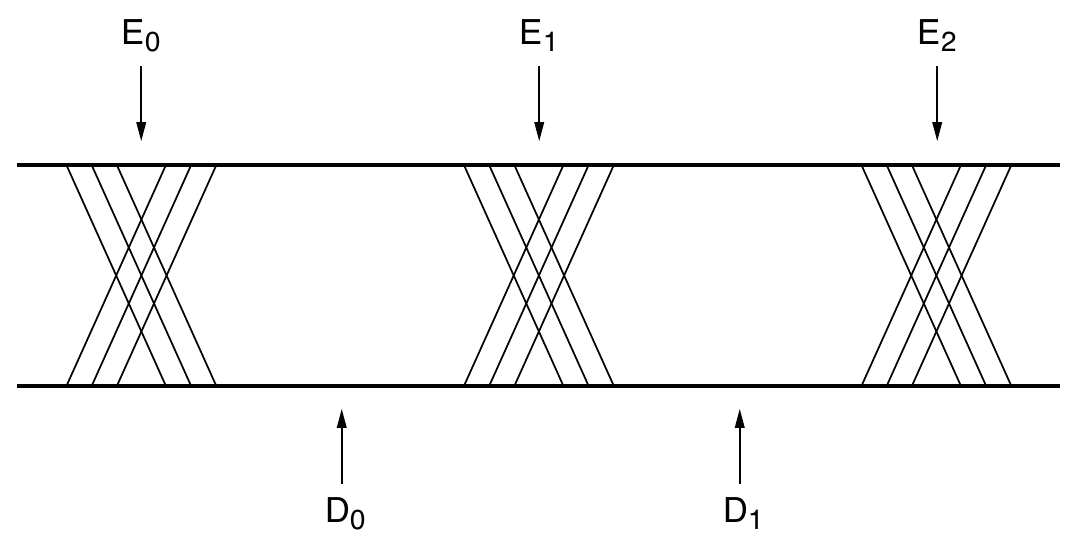}
	\vspace{-2mm}
	\caption{The CDR sampler\cite{bibGTXGTH}}\label{cdr}
\end{wrapfigure}

On one FELIX card, there are 24 to 40 bidirectional GBT links connected to the front-end. For this multi-channel project, the limit of the clock resources in FPGA should be considered. The FPGA on the FELIX prototype is Kintex Ultrascale\,\cite{bibTWEPP2016} which has more than 1000 global buffers, so each GBT channel can use independent clocks. While for the Virtex-7 FPGA on the FELIX demonstrator, the number of global buffers is 32, so the clocks must be shared between different channels. Taking the project in Virtex-7 FPGA as an example, the system clock distribution is shown in Figure\,\ref{clk_dist}. The transceiver reference clock GTHRefClk can originate from the dedicated reference clock pins MGTRefClk, or from the internal reference clock GRefClk. Though the GRefClk is not recommended by Xilinx, tests on several 7 series FPGAs showed that it works well when a good quality quad PLL is used by the transceiver. At the TX side, the 240 MHz GBT encoding clock TxWordClk can be driven by the TxOutClk from the transmitter, or by a 240 MHz GBTTxClk from another firmware block. When the latter one is used, its phase should be adjustable, to make sure no error occurs when the 20-bit GBT data is transferred to the TxOutClk domain as shown in Figure\,\ref{changes}. To save clock resources at the receiver side, the 240 MHz GBT decoding clock RxWordClk and the 40 MHz GBT frame clock RxFrameClk are shared among all of the channels. One channel is chosen as the master channel. The common RxFrameClk can be driven by the recovered clock RxOutClk from this master channel or the GBTRxClk from another firmware block. The latter one can be used only when it is synchronized with the recovered RxOutClk. When RxOutClk from the master channel is used, the clock domain crossing from RxOutClk to the RxFrameClk needs to be considered for all slave channels, since the RxOutClk of each channel are synchronized but have different phases. If the local GBTRxClk is used, this domain crossing should be solved for all channels. 

\begin{wrapfigure}[9]{r}{0.47\textwidth}
	\vspace{-3mm}
	\centering
	\includegraphics[width=0.47\textwidth]{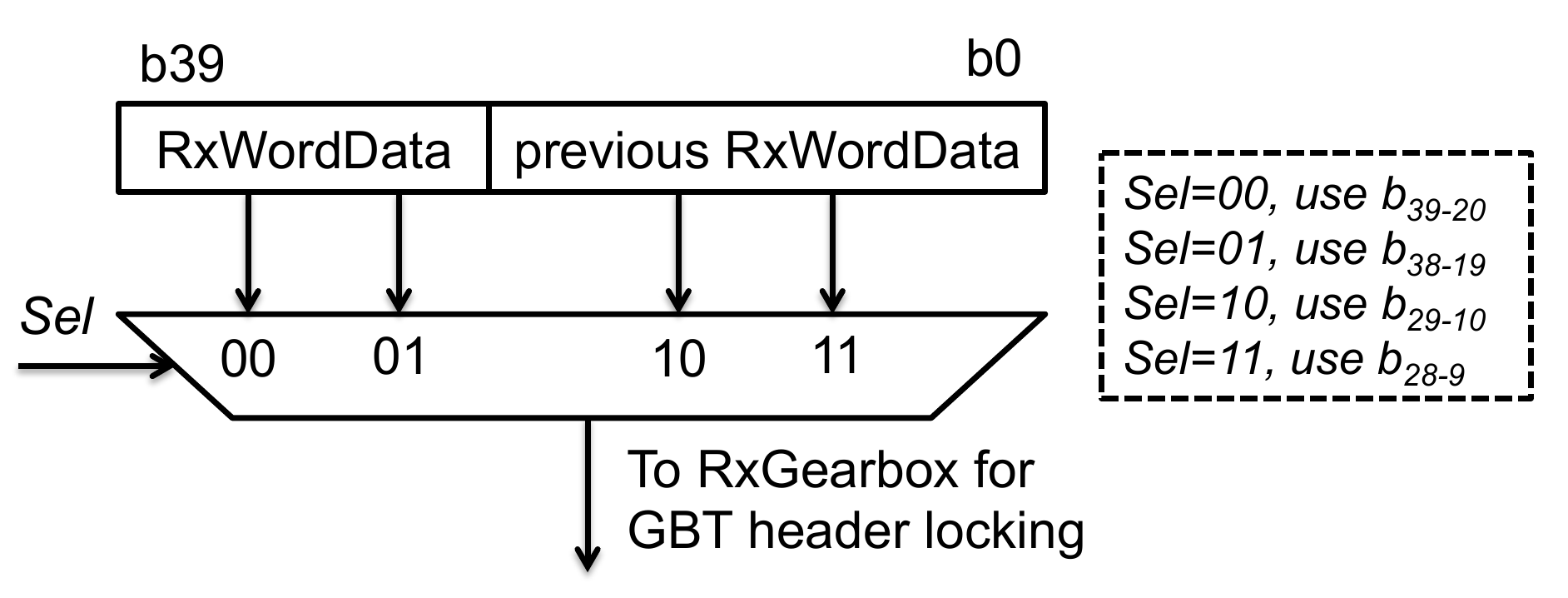}
	\vspace{-2mm}
	\caption{The multiplexer before RxGearBox}\label{rgbmux}
\end{wrapfigure}

The Figure\,\ref{cdr} shows the edge sampling and data sampling of the CDR (Clock and Data Recovery) in the Xilinx transceiver\,\cite{bibGTXGTH}. The FSM in the CDR will shift the phase of the recovered 2.4 GHz clock to make its edges being aligned to the center of eye-diagram. Since the data is DDR, means the recovered clock has two possible phases, the rising edge could be at $D_{0}$ or $D_{1}$. Inside the receiver, the 240 MHz RxOutClk is generated from this recovered 2.4 GHz clock. A 20-bit data is also created from ten cycles of the 2-bit data in the 2.4 GHz domain. The Xilinx transceiver provides a bitslip function which supports to shift the 240 MHz RxOutClk in a step of 1/2.4GHz. When shifting the clock by 1 step, the 20-bit data will also be shifted by 2 bits.  
So the 4-bit GBT header can either be shifted to bit 3-0 of the 20-bit data output (RxWordData in Figure\,\ref{rxtd}) of the receiver, or to bit 2-0 and the MSB (Most Significant Bit) of the last 20-bit output. Our testing shows that if the reference clocks at TX and RX side are not synchronized, the 2.4 GHz recovered clock can be shifted by 1/4.8GHz, by resetting the receiver\,\cite{bibODD}. As the shift step of the 240 MHz clock RxOutClk is 1 UI, and the 20-bit data shift step is 1 bit, the GBT header can always be shifted to the bit 3-0. Each time after the power cycling or firmware reset, the RX\_FSM will do this header locking automatically. When it is done, the phase of the recovered 240 MHz RxOutClk is fixed, that also means the latency from TX side to RX side is fixed. When these two reference clocks are synchronized, and if edges of the 2.4 GHz CDR reference clock at RX side are close to the edges of the eye diagram, a reset of the receiver can also make the recovered 2.4 GHz clock to be changeable between at $D_{0}$ or $D_{1}$. If the edges of the CDR reference clock are close to the middle of the input eye diagram, the receiver reset cannot shift the 2.4 GHz recovered clock between $D_{0}$ and $D_{1}$. At this time the latency is still fixed, but the 4-bit GBT header could be at bit 2-0 and the MSB of the last 20-bit output. For this case, an external 1 bit shift should be implemented to shift the GBT header to bit 3-0. To realize it bit 0 of the Sel signal shown in Figure\,\ref{rgbmux} should be '1', to choose bit 38-19 or bit 28-9 as the RxGearbox input. Another method is introduced in the reference\,\cite{bibMMCM}, an external MMCM (Mixed-Mode Clock Manager) is used to shift the clock by 1 UI\,\cite{bibMMCM}, with this method no receiver reset is required.


For the slave channels, concerning the clock domain crossing from the RxOutClk to the RxWordClk as in Figure\,\ref{changes}, the phase between them may be at any point of 0 to 360 degree. Enabling the RX buffer inside the transceiver is a method to guarantee no error occurs, but this increases the latency, and may cause latency uncertainties. As shown in Figure\,\ref{rgbmux}, a multiplexer is added before the RxGearbox. By changing the bit 1 of Sel, automatically bitslip operations controlled by the RX\_FSM in Figure\,\ref{changes} will shift the 20-bit data RxWordData from the transceiver by 10 bits, and the recovered RxOutClk by 10 UI, or half a Cycle. With this function there will always be at least one multiplexer selection which makes the data transmission from the RxOutClk to the common RxWordClk being stable. In the RX\_FSM, the RxWordClk is used to sample the RxOutClk. By repeating the bitslip for 10 times, a 10-bit result can be acquired for the 10 RxOutClk steps. With this result, the phase between these two clocks can be calculated, the bit 1 of Sel value can be determined. All of these operations are automatically done by the module RX\_FSM in Figure\,\ref{changes}. The clock jitter may change the measured 10-bit result, and the calculated Sel value. This doesn't cause bit error to the data transmission, but may change the 20-bit data in the RxWordClk be shifted by one Cycle. To solve it, when the system is finalized, a database can be used to store the measured Sel value and load it for future using. When any part of the TX or RX side is changed, the database should be updated.
    
\section{Conclusion}

The GBT link is used between FELIX and the front-end electronics system in the ATLAS upgrade framework. For the ATLAS Phase-I upgrade, the fixed low-latency is a basic requirement for the down-links from FELIX to the front-ends. Some optimizations are applied to the official GBT-FPGA core to decrease the latency. With this modified GBT-FPGA core, the GBT mode for TX and RX sides can be changed separately without FPGA reprogramming. The optimized core has been used on a KC705 evaluation board, the FELIX card and the gFEX board of the L1Calo system\,\cite{bibgFEX}. The fixed latency from the TX side 40 MHz TTC clock to the RX side 40 MHz TTC clock can be reduced by about 2 BC. The total latency for the two GBT modes is decreased to 3$\sim$3.5 BC. The multi-channel low latency design has been used on the GBT links between FELIX card and the gFEX board. The integration test shows the clock sharing scheme described above works well. The optimizations presented in this paper do not rely on the FPGA architecture, they can be implemented in different types of FPGAs.

\acknowledgments

The authors would like to thank for the original work of the CERN GBT-FPGA group. We also appreciate the help from Frans Schreuder and Andrea Borga at Nikhef.

\end{document}